\documentclass[11pt,a4paper]{article}
\usepackage{epsf,multicol,ifthen}
\usepackage[utf8]{inputenc}
\usepackage{amsmath}
\usepackage{amsfonts}
\usepackage{amssymb}
\usepackage{mathtools}
\usepackage{graphicx}
\usepackage{cite}
\usepackage[capitalise]{cleveref}
\usepackage{cite}
\usepackage{hyperref}
\usepackage{mathtools,amscd}
\graphicspath{ {./figures/} }
\usepackage[left=1cm,right=1cm,top=2cm,bottom=2cm]{geometry}
\crefformat{section}{#2section~#1#3}

\author{Tetiana Obikhod, Ievgenii Petrenko}
\title{The search for dark matter candidate, ALP, 
together with CP-odd Higgs boson and tau leptons at $\sqrt{s}=14$ TeV}
\date{%
    Institute for Nuclear Research NAS of Ukraine, Kyiv 03028, Ukraine\\%
    \today
}
\begin{document}

\maketitle

\abstract
{The cosmological observations of gravitational lenses, cosmic microwave
background, rotation speed of stars in galaxies confirm the existence of about 27\% dark
matter in the Universe. The nature of these particles
is unknown, however, there are theoretical models Beyond the Standard Model (BSM),
such as superstrings and D-branes, which predict new particles of type WIMPs, dilatons,
axions, etc. Experiments to search for such particles are being actively carried out both in
space and on modern accelerators, and unambiguous information regarding the type of
particles has not yet been identified.}

\section{Introduction}

One of the most popular and long-studied dark matter candidate 
\cite{1.,2.,3.} 
 is the axion-like particle (ALP). The only way to identify this 
 particle by its interaction constant has long
been studied by many collaborations such as Belle II, LEP, CDF, ATLAS, CMS and so on
\cite{4.}. The lack of information regarding the mass, type 
of particle (scalar or pseudoscalar) and since
the interaction of this particle is only gravitational, this 
leads to complications in both
experimental identification and theoretical interpretation. Thanks to modern Monte Carlo
programs, it is possible to model the properties of such particles to indicate optimal mass
regions, kinematic restrictions on momentum and pseudo-rapidity to facilitate a targeted
search for BSM physics.
To search for ALP, simplified models are usually used with a minimal set of new
candidate dark matter particles with one mediator, which, 
depending on the model, can be
a vector, axial-vector, scalar or pseudoscalar particle 
\cite{5.}. We used the reference model of two 
Higgs doublets (2HDM) plus a pseudoscalar mediator $a$ denoted 
as 2HDM+$a$ \cite{6.}. The model is described by 12 additional parameters:\\
$\bullet$ Masses of the CP-odd, CP-even and charged Higgs bosons 
($A$, $H$, $H^{\pm}$) are set to be equal to each other; \\
$\bullet$ Mixing angles $\alpha$ and $\beta$ are constrained to 
be $\cos \left( \beta - \alpha \right) = 0$, where $\alpha$ 
is the mixing angle between CP-even Higgs bosons $h$ and $H$, 
and $\tan\beta$ is the ratio of the vacuum expectation values (VEV) of the two Higgs doublets; \\
$\bullet$ The mass of the lightest Higgs boson $h$ is fixed 
to be $m_h \approx 125$ GeV; \\
$\bullet$ The mixing angle between the two pseudoscalars, $\theta$, 
is required to satisfy $\sin \theta = 0.35$ or  $\sin \theta = 0.7$. 

\begin{figure}[htbp]\centering
 \includegraphics[width=0.72\textwidth]{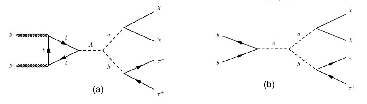}
 \caption{\label{fig:feyman} \small Feynman diagrams of the 2HDM+$a$ 
 model: (a) gluon-gluon merger and (b) $b$ $\overline{b}$ annihilation.}
\end{figure}

	The goal of our work is to use the MadGraph5\_aMC@NLO program 
\cite{7.} to model the production processes of the 
CP-odd Higgs boson, $A$, with subsequent decay into ALP 
and the lightest Higgs boson, $h$. So, the process of 
ALP formation together with the Higgs boson and two tau-leptons in the final state, 
$p p \to A \to a h \tau^+ \tau^-$, is studied. 
Representative Feynman diagrams for the dominant 
gluon-induced and quark-induced production and decay 
modes in the 2HDM+$a$ are presented in Figure \ref{fig:feyman}.  

\section{Calculations}
The corresponding experimental search for ALP was performed 
by the ATLAS collaboration at the LHC during
 2015-2018 at an integral luminosity of 
 139 fb$^{-1}$ in proton-proton collisions 
 at $\sqrt{s}=13$ TeV,
  \cite{8.}. Using the restrictions on 2HDM+$a$ model 
parameters obtained from experimental data, we selected 
two sets of parameters, which are listed in Table 1.
\begin{center}
{\it Table 1. Parameter scenarios for the 2HDM+$a$ model}\\
\begin{tabular}{|c|c|c|c|c|}
\hline 
Scenarios & $m_a$ & $m_A$ & $\tan \beta$ & $\sin\theta$ \\ 
\hline 
BP1 & 500 GeV & 1400 GeV & 1 & 0.35 \\ 
\hline 
BP2 & 150 GeV & 500 GeV & 2 & 0.7 \\ 
\hline 
\end{tabular} 
\end{center}

\begin{figure}[htbp]\centering
	 \includegraphics[width=0.44\textwidth]{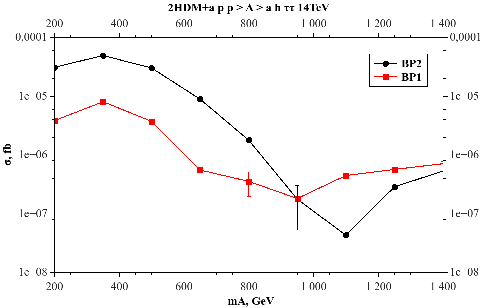}
	 \caption{\label{fig:cs_mA} Cross-sections for 
 Higgs boson formation, $A$ as a function 
         of Higgs boson mass, $m_A$.}
		 \includegraphics[width=0.44\textwidth]{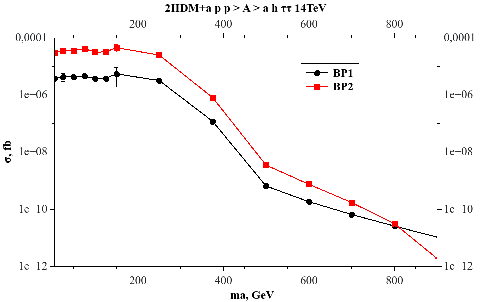}
		 	 \caption{\label{fig:cs_ma} Cross-sections 
 for axion formation as a function of the axion mass, $m_a$.}
 \end{figure}

	The use of the scenarios listed in Table 1 made it possible 
to calculate the cross-sections of the formation of bosons 
(Figure \ref{fig:cs_mA}) and axions (Figure \ref{fig:cs_ma}) 
using the MADGRAPH5@NLO computer program. 

\begin{figure}[htbp]\centering
 \includegraphics[height=0.2\textheight]{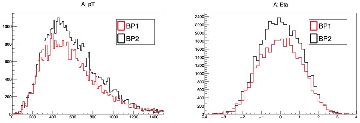}
 \includegraphics[height=0.2\textheight]{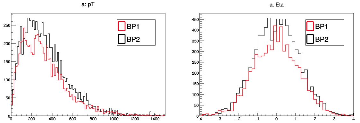}
 \caption{\label{fig:kinematics} Kinematical data for CP-odd bosons, A (up) and ALP (down).}
\end{figure}

The corresponding kinematic restrictions for momentum (left) 
and pseudorapidities (right) for CP-odd bosons, 
$A$, and ALP are shown in Fig. \ref{fig:kinematics} 
From the data presented in Fig. \ref{fig:kinematics}, 
it is necessary to conclude about the predominance of momentum values 
in the region of 400-600 GeV for CP-odd Higgs boson,A, and from 100 to 400 
GeV for ALP. Regarding pseudorapidity, 
for the CP-odd Higgs bosons, a larger number of 
events are located in the region from $-3$ to $3$. 
For ALP, the behavior of pseudorapidity is more symmetrical for BP2 scenario
in region from $-1.5$ to $1.5$.

\section{Conclusions}
The 2HDM+$a$ model for searching and modeling ALP particles 
is considered. Experimental constraints on the search for 
ALP particles became the basis for the selection of 
simulation scenarios. The obtained kinematic constraints 
on the transverse momentum and angular distribution allow 
to realize possible searches for the CP-odd Higgs boson 
with a mass in the region of 400 GeV and a dark matter 
candidate, ALP, with a mass up to 200 GeV. The obtained 
data demonstrate the priority of the second BP2 scenario, 
which correlates with the latest experimental constraints 
of the ATLAS collaboration.

\end{document}